\documentclass[a4paper]{article}

\usepackage{amsmath,amssymb}

\newcommand{\ii}{\mathrm{i}}

\newcommand{\dd}{\mathrm{d}}
\newcommand{\pd}{\partial}

\newcommand{\e}{\mathrm{e}}

\newcommand{\tr}{\mathop{\mathrm{tr}}\nolimits}

\begin{document}

\title{Continuum limits of Berenstein--Maldacena--Nastase matrix theory: Where is the
(nonabelian) gauge group?}

\author{\textbf{Corneliu Sochichiu}\thanks{
On leave from: \textit{Bogoliubov Laboratory of Theoretical
Physics, Joint Institute for Nuclear Research, 141980 Dubna,
Moscow Reg., RUSSIA} and \textit{Institutul de Fizic\u a Aplicat\u
a A\c S, str. Academiei, nr. 5, Chi\c sin\u au, MD2028 MOLDOVA};
e-mail: \texttt{sochichi@lnf.infn.ti}}\\
High Energy and Elementary Particle\\
Physics Division,
University of Crete,\\
P.O. Box 2208, GR-710 03 Heraclion,\\ Crete,
GREECE\\
and\\
Laboratori Nazionali di Frascati,\\
INFN, Via E. Fermi 40,\\
C.P. 13, 00044 Frascati\\
Italy
}

\maketitle
\begin{abstract}
We discuss the continuum limits of Berenstein-Maldacena-Nastase
matrix model. The special attention is paid to limits that give
rise to Poisson bracket gauge field theories with gauge groups
U($n$) on the ordinary two sphere. The gauge group and the space
depending on the degeneracy of the classical solution about which
the model is considered. We compare these limits as well as
different solutions in the framework of the same limit model. We
show that these models fail to be equivalent in the continuum
limit, i.e. the continuum limit does not commute with dualities of
the matrix theory.
\end{abstract}


\section{Introduction}
The correspondence between gauge fields and strings is a long
standing problem \cite{'tHooft:1974jz}. To establish such an
equivalence a good understanding of both gauge and string theories
is needed, inclusively on the nonperturbative level. Some progress
in this direction was reached recent years. (For a comprehensive
review see \cite{Aharony:1999ti}.)

On the other hand, hints for connection between nonperturbative
string dynamics and matrix models which are dimensionally reduced
Yang--Mills theories where given by matrix theories
\cite{Banks:1997vh,Ishibashi:1997xs}, describing branes.

In recent papers
\cite{Berenstein:2002jq}\nocite{Berenstein:2002zw}--\cite{Berenstein:2002sa},
the correspondence between Yang--Mills theory and string theory in
pp-wave background \cite{Blau:2002dy,Blau:2002mw}, was considered.
The pp-wave background is a curved space corresponding to a plane
parallel gravitational wave. In such a space the string theory
preserves many of its nice features, in particular it is treatable
and can be quantized \cite{Metsaev:2001bj}.

The DLCQ compactification which in the flat space gives the BFSS
model \cite{Banks:1997vh}, in the case of pp-wave leads to a
modified matrix theory for zero-brane. The modification with
respect to the BFSS model results in the addition of mass terms
for all matrix fields and of Chern--Simons terms to some of them.
The result of this modification is that the stationary vacua of
the model are given by fuzzy sphere or by a set of fuzzy spheres.
The perturbation theory around such classical vacua was analyzed
\cite{Dasgupta:2002hx} together with the continuum limit (see also
\cite{Valtancoli:2002rx}). Also, in \cite{Dasgupta:2002hx}, it was
shown that BMN matrix theory arises as the result of the world
sheet quantization of membranes in pp-wave background described by
a Poisson bracket model. Like BFSS matrix theory the BMN matrix
theory also possesses a number of supersymmetric solutions
\cite{Bak:2002rq}--\nocite{Bak:2002aq,Sugiyama:2002jq}\cite{Sugiyama:2002rs}.

Here we review the fact that in the scaling limit proposed by BMN
one in fact recovers the above Poisson bracket action of the
commutative spherical brane. Depending on what solution of the
matrix model is chosen, one can get in the above limit field
models with different local gauge groups. In earlier papers
\cite{Sochichiu:2000ud}\nocite{Sochichiu:2000bg,Sochichiu:2000kz,%
Kiritsis:2002py}--\cite{Sochichiu:2002jh} we analyzed the
equivalence relations arising in the $N\to\infty$ limit of
``flat'' matrix theories. In that cases, $N\to\infty$ limit
yielded noncommutative gauge models in different dimensions and/or
having different gauge groups. Now, the situation is slightly
different. Since the $N\to\infty$ limit of BMN matrix theory is
combined also with the commutative limit, the scaling prescription
leads to a commutative, although exotic, field model on an
ordinary sphere or set of spheres. The equivalence of these
different limits in this case is not clear \emph{a priori}. To
find such an equivalence, if it exists, one has to identify the
solutions in the limit model, which correspond to different sets
of spheres and analyze the model in the vicinity of such
solutions. We do this and find that, in fact, the models are not
equivalent. One can recover, starting with the U(1) model, the
spectrum corresponding to the Abelian subalgebra only of
nonabelian models.

The plan of the paper is as follows. In the next section we review
the classical solutions and continuum limit of the BMN matrix
model. After that, we consider the Poisson gauge model with the
gauge group U(1) and find in this model solutions which are
commutative analogs of U($n$) backgrounds of matrix theory. We
find that it is only maximal abelian subgroup U(1)$^n$ of U($n$)
which is manifest while the remaining part is hidden in the large
gauge transformations of the original irreducible limit. The world
sheet quantization makes these modes to arise explicitly.

\section{Classical solutions and continuum limit}
The BMN matrix model appears as the DLCQ quantization of zero
brane in the pp-wave background,
\begin{equation}\label{pp}
ds^2=-4\dd x^+\dd
x^--\left[\left(\frac{\mu}{3}\right)^2x_\alpha^2+
\left(\frac{\mu}{6}\right)^2x_\mu^2\right](\dd x^+)^2+\dd x_i^2,
\end{equation}
were the early Greek indices run $\alpha,\beta=1,2,3$, late ones
$\mu,\nu=4,\dots ,9$, while the Latin ones span both of these sets,
i.e. $i=1,2,\dots,9$, $x_i=(x_\alpha,x_\mu)$. In this approach,
the sector of M-theory corresponding to the light cone momentum
$2p^+=-p_-=N/R$ ($R$ is the DLCQ radius) is described by the following
matrix action \cite{Berenstein:2002zw},
\begin{multline}\label{S}
S=\int \dd t \tr \left[\frac{1}{2(2R)}(D_0\phi^i)^2
+\frac{(2R)}{4}[\phi^i,\phi^j]^2-\right. \\
\left.-\frac{1}{2(2R)}\left(\left(\frac{\mu}{3}\right)^2\phi_\alpha^2+
\left(\frac{\mu}{6}\right)^2\phi_\mu^2\right)-\frac{\mu}{3}\ii
\epsilon_{\alpha\beta\gamma} \phi_\alpha\phi_\beta\phi_\gamma\right]
+\text{fermions},
\end{multline}
where $\phi_i$ are $N\times N$ hermitian matrices and ``fermions''
denotes the fermionic part of the action which is not written
explicitly, since it is not important for our further analysis. In
\eqref{S} the indices run according to the same convention as in
eq. \eqref{pp}.

One can see, that there are no stable nontrivial vacua involving
only fields $\phi_\mu$ (cfy. Ref. \cite{Valtancoli:2002rx}), while
one can build nontrivial vacuum solutions out of $\phi_\alpha$. In
what follows, we will consider the model about such
configurations.

The $\phi_\alpha$ dependent part of the action can be rewritten in
the following form:
\begin{equation}\label{su2ac}
S_0=\int\dd t\tr\left[\frac{1}{2(2R)}(D_0\phi_\alpha)^2+
\frac{(2R)}{4}\left([\phi_\alpha,\phi_\beta]-
\ii\frac{\mu}{6R}\epsilon_{\alpha\beta\gamma}\phi_\gamma\right)^2\right].
\end{equation}

As it is not difficult to see from the form \eqref{su2ac} of the
action, the vacua of this sector of the model are given by
matrices satisfying $su(2)$ algebra,
\begin{equation}\label{su2}
[\phi_\alpha,\phi_\beta]=
\ii\frac{\mu}{6R}\epsilon_{\alpha\beta\gamma}\phi_\gamma.
\end{equation}

The matrices $\phi_\alpha$, satisfying vacuum condition
\eqref{su2} can be split into blocks corresponding to irreducible
representations $R_\lambda$ of $su(2)$ of spins $j_\lambda$,
having the total dimensionality,
\begin{equation}\label{sum-rep}
\sum_\lambda (2j_\lambda+1)=N.
\end{equation}

The cases of interest for us are when the solution is represented
by $n$-times degenerate irreducible representation of the spin $j$
satisfying $2j=N/n-1$. In particular, the simplest case is for
$n=1$ when one has a simple irreducible representation with
$2j+1=N$. Although there are other interesting cases, in what
follows we concentrate mainly on the above ones.

So, let us consider a solution $\phi_\alpha\equiv Y_\alpha$, which
is $n$ times irreducible representation. An arbitrary Hermitian
$n\times n$ matrix can be uniquely expanded in terms of $n\times
n$ Hermitian matrices whose entries of symmetrised traceless
polynomials of $Y_\alpha$. This polynomials are noncommutative
analogues of spherical functions and treating them as such one has
a map from the space of operators on $N$ dimensional space to the
space of $n\times n$ matrix valued functions on a sphere of the
radius,
\begin{equation}\label{radius}
Y_\alpha^2= \left(\frac{\mu}{6R}\right)^2j(j+1).
\end{equation}
These functions are subject to the star product on fuzzy
sphere whose exact form we will not need\footnote{For the details
referring the fuzzy sphere star product we
send readers to \cite{Valtancoli:2002rx,Dasgupta:2002hx}.}.

Thus, an arbitrary matrix configuration can be considered as a
perturbation of the background solution,
\begin{subequations}\label{exptn}
\begin{align}
& \phi_\alpha=Y_\alpha+A_\alpha, \\
& \phi_\mu=\phi_\mu,\\
& \dots,
\end{align}
\end{subequations}
where $A_\alpha$ and $\phi_\mu$ are now fields on the fuzzy sphere
and dots stay for fermions. In this parametrization the action
\eqref{su2ac} is essentially one of Yang--Mills--Higgs model on a
fuzzy sphere.

Having in mind this map one can switch between different solutions and
re-expand as in \eqref{exptn} to obtain equivalence maps between
models with different gauge groups living on fuzzy spheres of
different radii which are related as $rn=$const (see
\cite{Sochichiu:2000ud}\nocite{Sochichiu:2000bg,Sochichiu:2000kz,%
Kiritsis:2002py}--\cite{Sochichiu:2002jh}).

According to BMN prescription, as $N$ goes to infinity the radius of
the sphere remans finite,
\begin{equation}\label{N->infty}
Y_\alpha^2\to r_0^2=\left(\frac{\mu p^+}{6n}\right)^2,\qquad
N\to\infty, \qquad (2R)\sim \frac{N}{p^+},
\end{equation}
while the background becomes commutative,
\begin{equation}
[Y_\alpha,Y_\beta]=\ii
\frac{n}{N}r_0\epsilon_{\alpha\beta\gamma}Y_\gamma\to 0.
\end{equation}
Ploughing this into action one should be careful with divergent
factors of $(2R)$. The contribution to the action will be given by
the leading term in the expansion of commutators,
\begin{equation}\label{com->pb}
[f,g]\approx \ii\frac{n}{N}\{f,g\}+0(N^{-2}),
\end{equation}
where $\{,\}$ is the Poisson bracket on the sphere which is given by,
\begin{equation}\label{pb}
\{f,g\}=\epsilon_{\alpha\beta\gamma}Y_\alpha\pd_\beta f\pd_\gamma g.
\end{equation}
Also the trace is replaced by integration over the sphere according
to,
\begin{equation}\label{tr->int}
\frac{4\pi n}{N}\tr\to \int \dd \Omega.
\end{equation}

Then, the action becomes,
\begin{multline}\label{N->infty-ac}
S_{N\to\infty}=\int\dd t\dd\Omega \tr_{u(n)}\left(\frac12
(D_0\phi_i)^2+
\frac{1}{4g^2}(\{\phi_\alpha,\phi_\beta\}-
r\epsilon_{\alpha\beta\gamma}\phi_\gamma)^2\right.\\
\left.+\frac{1}{4g^2}\{\phi_\mu,\phi_\nu\}^2+
\frac12\left(\frac{r}{2g}\right)^2\phi_\mu^2\right)+\text{fermions},
\end{multline}
where $\phi_\alpha =Y_\alpha+A_\alpha$, $g^2=(p^+)^2$. Integration
is performed over time and a sphere whose radius is $r=(\mu
p^+)/6n$, all fields are $n$ dimensional Hermitian matrices
subject to $u(n)$ trace.

The model possesses a Poisson bracket gauge symmetry,
\begin{equation}\label{pb-gt}
\phi_i\to\phi_i+\{\phi_i,u\},
\end{equation}
where $u$ is an arbitrary hermitian $n\times n$ matrix valued
function.

\section{Commutative ``dualities''}
In the previous section we found that the $N\to\infty$ limit of
the BMN matrix theory is sensitive to the background around which
we are considering it. For any finite $N$ and finite
noncommutativity they are just different parameterizations of the
same matrix model and, therefore, these models are all equivalent.
This may not hold true as $N$ goes to infinity and the
noncommutativity vanishes. Let us try to check, however, at which
extend this equivalence is still present in the limiting model
\eqref{N->infty-ac}.

In order to do this consider the model \eqref{N->infty-ac} for
$n=1$ and $r=r_0$ which is obtained from the $N\to\infty$ limit of
the irreducible algebra. The gauge symmetry here is just U(1)
Poisson bracket gauge symmetry. Let us find static vacuum
solutions of this model most close to one producing the U($n$)
model.

The static vacua satisfy an equation analogous to \eqref{su2} that
the commutator is replaced with the Poisson bracket,
\begin{equation}\label{pbem}
\{\phi_\alpha,\phi_\beta\}=r_0\epsilon_{\alpha\beta\gamma}\phi_\gamma.
\end{equation}
$\phi_\alpha$ are functions on (ordinary) sphere of radius
$r_0$. Since $\phi_\alpha^2$ P.b.-commutes with all $\phi_\alpha$ the
solution is, in fact a map of two spheres: $S^2\to S^2$. Nontrivial
solutions are given, therefore, by the homotopically nontrivial
maps. Since $\pi_2(S^2)=\mathbb{Z}$ it is natural to identify the
homotopy class of the solution with the rank $n$ of the gauge group U($n$). Indeed,
zero and one class solutions correspond to $\phi_\alpha=0$ and
$\phi_\alpha=Y_\alpha$, respectively.

Let us find higher classes. In spherical coordinates,
\begin{subequations}\label{sph}
\begin{align}
& Y_1=r_0\sin\theta\cos\varphi,\\
& Y_2=r_0\sin\theta\sin\varphi,\\
& Y_3=r_0\cos\theta,
\end{align}
\end{subequations}
the Poisson bracket is given by,
\begin{equation}
\{f,g\}=\frac{1}{\sin\theta}(\pd_\theta f\pd_\varphi g-
\pd_\theta g\pd_\varphi f).
\end{equation}

On the other hand, the simplest way to get a map of $n$-th
homotopically class is to wrap along $\varphi$,
\begin{subequations}\label{n-sph}
\begin{align}
& \phi_1\equiv Y_1^{(n)}=r\sin\theta\cos n\varphi,\\
& \phi_2\equiv Y_2^{(n)}=r\sin\theta\sin n\varphi,\\
& \phi_3\equiv Y_3^{(n)}=Y_3=\cos\theta.
\end{align}
\end{subequations}

Fortunately, we are lucky enough and the map as it appears in
\eqref{n-sph}  satisfies the vacuum condition \eqref{pbem} if the
radius $r$ is chosen to be $r=r_0/n$. This relation is encouraging
since it is exactly the relation of the radii of the spheres on which
U($n$) models live in the $N\to\infty$ limit (cfy. \eqref{N->infty}).

To proceed further we have to consider the functions \eqref{n-sph}
as new ``coordinates'' by which we should substitute the old ones.
Since $Y^{(1)}$ are wrapping $n$ times about $Y^{(n)}$ a generic
function of $Y^{(1)}$ becomes a $n$-fold ambiguous function of
$Y^{(n)}$. Locally, any function of $Y^{(1)}$ will become a set of
$n$ functions of $Y^{(n)}$,
\begin{equation}
\phi(Y^{(1)})\mapsto \phi^a(Y^{(n)}),\qquad a=1,\dots,n,
\end{equation}
one for each sheet. (In general functions are mapped to sections of a
nontrivial $n$ dimensional fibre bundle.)

Unfortunately, this is not in a total accordance with our
expectations, since in order to get U($n$) gauge group the fields
should map to $n\times n$ dimensional matrices rather then to $n$
component fields. In fact, the fields in new coordinates represent
the diagonal part of the expected matrices. Indeed, the gauge
transformation \eqref{pb-gt} splits in $n$ U(1) parts (one U(1)
for each sheet) which is an indication that the gauge group is
U(1)$^n$.

Summarizing, it appears that the maximum we can get in the
limiting model is to map the U(1) model to a model were U($n$) is
truncated down to U(1)$^n$.

\section{World volume quantization and restoration of the whole U($n$)
group}

Let us try to recover the remaining non-diagonal part of the
desired U($n$) symmetry group.

In fact, the symmetry can be restored upon the worldvolume
quantization. The idea can be illustrated by the following
example. Consider a particle moving in a space consisting of $n$
sheets (branes). The position of the particle is given by its
(continuous) coordinate $x$ and the number of the sheet
$a=1,\dots,n$. Classically, the particle can move smoothly along
$x$ and jump through indices $a$. Suppose the observer does not
care about the sheet numbers. So far there is no nonabelian
symmetry in the system.

Now, consider the above model as being quantum. Since the particle
can be found on different branes, the wave function of the
particle is a $n$ dimensional vector $\psi_a(x)$. There are $n^2$
Hermitian operators describing the jumps of the particle from $a$
to $b$, which commute with $x$ and $p$ and which generate a $U(n)$
symmetry group.

Let us return now to our model. The gauge symmetry \eqref{pb-gt} is in
fact only the infinitesimal version of the whole gauge
invariance. Eq. \eqref{pb-gt} can be integrated to Hamiltonian flows
to yield the finite gauge transformations. For example, a rotation by
$\Delta\varphi$ along the $Y_3$ axis can be formally written as,
\begin{equation}
\phi(\varphi)\mapsto\phi(\varphi+\Delta\varphi)=\e^{\{\Delta\varphi
Y_3,}\phi\}(\varphi),
\end{equation}
where we symbolically denoted the exponentiated Poisson bracket,
\begin{equation}
\e^{\{A,}B\}=B+\frac{1}{1!}\{A,B\}+\frac{1}{2!}\{A,\{A,B\}\}+\dots
\end{equation}

Thus the ``large'' rotations of $\varphi$ by, say, $2\pi k$ where
$k<n$ is an integer, result in cyclic jumping over $k$ sheets.
$Y^{(n)}$ are invariant under such transformations since they are
degenerate along the sheet numbers. The total number of
independent large rotations is precisely $n^2$ (including
identical rotations). Thus the nonabelian structure is hidden in
large gauge transformations!

Consider now the world volume quantization. It results in the
replacement of the Poisson bracket algebra,
\begin{equation}
\{Y_\alpha^{(n)},Y_\beta^{(n)}\}=r
\epsilon_{\alpha\beta\gamma},Y_\gamma^{(n)},
\end{equation}
by an operator one,
\begin{equation}\label{qsph}
[Y_\alpha^{(n)},Y_\beta^{(n)}]=\ii \hbar r
\epsilon_{\alpha\beta\gamma},Y_\gamma^{(n)},
\end{equation}
in such a way that it forms an irreducible representation modulo the
action of the sheet jump operators. It is $n$-tuple degenerate one and
this degeneracy is governed by U($n$) gauge group. Now arbitrary
operator about the background $Y^{(n)}$ is represented by
$n$-dimensional matrix valued noncommutative function on the fuzzy
sphere \eqref{qsph}.

Let us note that the gauge group was restored at the moment when we
replaced the ``classical'' sheet number label by an operator.

\section{Discussion}
In this note we considered the properties of $N\to\infty$ limit of
the BMN matrix theory.

Field theory models describing the fluctuations of the matrix
theory in $N\to\infty$ limit depend on the classical background
around which the fluctuations are measured. Different backgrounds
lead to models in different spaces or having different gauge
groups. In BMN matrix theory there is a class of classical
backgrounds given by fields satisfying $su(2)$ algebra. In the
case when such a background is given by a $n$ times degenerate
irreducible representation of spin $j\sim N/n$, the continuum
limit of the matrix model is described by a Poisson gauge model
with group U$(n)$. Since at finite $N$ the models with different
gauge groups appear as different parameterizations of the same
matrix theory there is an equivalence between them. As $N$ goes to
infinity, however, the models become nonequivalent. Thus, we are
able to map the Poisson gauge model with group U(1) to at most the
abelian sector of the U$(n)$ model. Quantization of the worldsheet
seems to restore the whole gauge group.

The string interpretation of the above properties of the limit is
as follows. Continuum limit of the BMN matrix theory corresponds
to zero slope or infinite tension limit of the string theory. So,
the transition from a one-brane configuration, which corresponds
to gauge group U(1), to a multi-brane configuration, corresponding
to group U$(n)$, passes through intermediary configurations with
concentric branes of different radii. Infinite tension strings can
not stretch between branes of different radii, but rather can
begin and end on the same brane. The last gives modes in the
abelian sector of the gauge theory, while nonabelian modes can not
appear even when spherical branes become of the same radius.

Here we considered only very specific backgrounds given by
irreducible representations or product of identical irreducible
representations. Another interesting class is given by background
with products of \emph{different} irreducible representations. In
the case of these solutions one can expect to have a Poisson gauge
model with spontaneously broken gauge group U$(n)$ $\to$
U$(n_1)\times\dots\times$U$(n_k)$, $n_1+\dots+n_k=n$, where $n_i$
is the multiplicity of the $i$-th irreducible factor. These can
serve as subjects for a future study.

Unfortunately, there is no known direct relation between BMN
matrix theory and BMN super--Yang--Mills/PP (SYM/PP) string
correspondence \cite{Berenstein:2002jq}, except that the
respective matrix theory was designed to describe the brane
dynamics on the pp-wave. It would be interesting to know such a
relation in case it exists, but so far the SYM/PP correspondence
is formulated in terms of perturbative string states and extending
the analysis beyond the perturbative limits is a good challenge.

\section*{Acknowledgments}
I benefited from useful discussions with K.~Selivanov, R.~Woodard,
K.~Anagnostopoulos. I would like to thank Th.Tomaras for kind
hospitality and interest in my work.

This work was done during my visit to University of Crete
supported by Greek Nato fellowship program. I would like to thank
Mr. N. Sinanis of University of Crete for various help during my
stay in Crete.

This work was also supported by RFFI grant
02-01-00126a and the grant for support of leading scientific schools
0015-96046 and INTAS grant 000262.

\providecommand{\href}[2]{#2}\begingroup\raggedright\endgroup

\end{document}